\documentclass[12pt]{iopart}

\usepackage{graphicx}
\usepackage{xcolor}

\begin{document}



\title{Multiphysics Simulation and First Prototype Development of a Microwave Plasma System for Chemical Vapour Deposition (CVD) Applications}


\author{ Akash Akash$^{1}$, Sanjeev Kumar Pandey$^{1}$, Venkata Sai Teja Madana$^{1}$, S. Manikandan$^{1}$, Guhan Gunasekaran$^{1}$, Sundar Ravi$^{1}$, Sethu Narayanana S$^{1}$, C. Nikhil$^{1}$, Nishant Sirse$^{1,4}$, S. Sathyan$^{1,3}$, N. Arunachalam$^{1,3}$, and M.S Ramachandra Rao$^{1,2,3}$}

\address{$^{1}$India Center for lab-grown Diamond (InCent-LGD), Indian Institute of Technology Madras, Chennai, India}
\address{$^{2}$Department of Physics, Materials Science and Research Centre (MSRC), Nano-Functional Materials Technology Centre (NFMTC), 
  Quantum Centre of Excellence for Diamond and Emergent Materials (QuCenDiEM), Indian institute of Technology Madras, Chennai, India}
\address{$^{3}$Department of Mechanical Engineering, Indian institute of Technology Madras, Chennai, India}
\address{$^{4}$Institute of Science and Research, and Centre for Scientific and Applied Research, IPS Academy, Indore, India}  
\ead{akash@incentlgdiitm.org}
\vspace{10pt}
\begin{indented}
\item[]August 2025
\end{indented}

\begin{abstract}
With the aid of COMSOL multiphysics simulations, a compact microwave plasma reactor operated at 2.45 GHz frequency has been designed for diamond film deposition. The reactor consists of a cylindrical cavity that resonates in the fundamental mode $TM_{01p}$ with a longitudinal field variation (p = 1). Investigations on microwave electric field and hydrogen $(H_{2})$ plasma characteristics inside the microwave plasma cavity have been carried out, which assisted in the resonant cavity optimizations. The new reactor design includes a unique antenna structure which facilitates better thermal management and gas inlet arrangement. Parametric analysis of the effect of increase in microwave power, gas pressure and synergistic effects  of power and pressure variations on the $H_{2}$ plasma characteristics such as electron density, gas temperature, and atomic hydrogen density  have been performed computationally to estimate the optimize reactor operating conditions. Observations from our simulations indicate that the cavity design is able to operate within a range of microwave power and gas pressure upto $P_{in}=6$ kW and $p_{0}=30$ kPa respectively. Preliminary experimental validation which includes vacuum integrity and $H_{2}$ plasma ignition tests inside the cavity are also reported.       
\end{abstract}
%
\vspace{2pc}
\noindent{\it Keywords} : Microwave plasma chemical vapour deposition (MPCVD), Low-temperature plasmas, Hydrogen plasma discharge, Fluid discharge modeling, Maxwell's equations, Drift-Diffusion model, Navier-Stokes equation, Plasma chemistry, COMSOL multiphysics simulations.
%
%
%
\ioptwocol
%

\section{Introduction}
\label{Introduction}

Diamond being a crystalline form of pure carbon, it possesses an extensive combination of physical, mechanical, and chemical properties. It has enormous potential for industrial applications such as optics \cite{POPOV_2006,Kolodziej_2016}, high-power electronics \cite{Graebner_1992,Goodson_1995,Angadi_2006,ANAYA_2016,LIU_2017} and other sectors \cite{MURAKAWA_1991,FJ_2005,AUCIELL_2010,Rath_2015,Nebel_2023}. There are two main approaches for diamond crystal synthesis, high-pressure high-temperature (HPHT) and chemical vapor deposition (CVD). Since its inception, the CVD method has become one of the most prominent methods for growing high quality diamond films \cite{Butler_2009,SCHWANDER_2011,ANGUS_2014}. Among various CVD techniques, microwave plasma chemical vapor deposition (MPCVD) has attracted more attention due to its following advantages,: absence of electrode contamination, possibility of control optimization to produce high-quality diamond, eco-friendly and ability to produce large area films due to high-density plasmas. However, the only disadvantage of the MPCVD technique is the slow growth rate at which diamond is deposited \cite{YF_LI_2014}. It is well known that for the enhanced growth rate of diamond deposition, the most efficient way is to increase the microwave power density within the MPCVD reactor cavities \cite{DING_2012,ZUO_2008,GU_2012,HEMAWAN_2010,YAMADA_2008,WENG_2012,YAMADA_2006}. In the past, MPCVD reactors have been evolved from tubular type \cite{KAM_1983}, quartz bell jar type \cite{BACHMANN_1991}, cylindrical cavity type \cite{Gu_2011,Collin_thesis_2013,Bai_2022}, to the recent ellipsoidal cavity type \cite{Funer_1998}, multi-mode non-cylindrical cavity, plate-to-plate cavity types \cite{Z_MA_2016} and CAP type reactor \cite{PLEULER_2002}. Also commercially, there are various MPCVD reactors with different transverse magnetic (TM) topologies are available in the market which include the $TM_{0(n>1)}$ type cavity such as the ARDIS-100, Carat Systems CTS6U, Seki Diamond SDS 6K style clamshell \cite{WENG_2018,Yamada_2012,YF_LI_2014,YAMADA_2015,SEDOV_2020}, cylindrical $TM_{01(p>1)}$ type cavity such as the ASTEX PDS-18, Seki Diamond SDS 5200 series reactors \cite{FUNER_1999,GORBACHEV_2001,YAMADA_2006,SILVA_2010,Shivkumar_2016}, the ellipsoidal egg-shaped cavity such as the AIXTRON reactor \cite{FUNER_1999,LI_2011} and the $TM_{02}$ dome style cavity \cite{SU_2014}. Keeping high microwave power density requirement for large deposition rates in mind, designing of the resonant cavity is the most non-trivial component one has to take care of while developing a MPCVD reactor.          

Computational modeling of the microwave plasma discharges inside the MPCVD reactors have become an essential tool for the design, optimization and development of these reactors \cite{Hasouni_1999,Hagelaar_2004,Hassouni_2010,LI_2011,LI_2017,Yan_2021,Mesbahi_2013,Prasanna_2016,ZHANG_2022,CUENCA_2022,YANG_2023}. Typically, the exact modeling of these plasma discharges requires a self-consistent multiphysics coupled approach which includes an electromagnetic (EM) solver for microwave EM fields, a plasma solver for transport and gas discharge chemistry, a thermal solver for energy estimation and temperature distributions, and a surface kinetics solver for deposition rate estimations. In the recent past, several successful attempts were made for designing the resonant microwave cavities as well as modeling these microwave plasmas \cite{Hasouni_1999,Tachibana_1984,Hasouni_1998,TAN_1994,Tahara_1995,FUNER_1999,HASSOUNI_1997,Bou_1992,TAN_1995,GORBACHEV_2001,Lombardi_2005,YAMADA_2007,Mankelevich_2008,Yamada_2011,Yamada_2012,Richley_2011,YAMADA_2015,ASHFOLD_2023}. From reactor designing point of view, authors \cite{Silva_Hasouni_2009} summarized standard operating principles, design methodology of a 2.45 GHz MPCVD reactor and its scalability to lower frequencies (915 MHz) for large area crystal growth process using hing density plasmas. Instead of resolving all the numerical complexities involved in these low temperature microwave discharge plasmas, several works have been reported in which plasma simulation had been carried out using reduced and phenomenological models \cite{LI_2011,SU_2014,LI_2014,Shivkumar_2016,Wang_2019}. The main advantage of these models are that it qualitatively predicts the plasma behavior inside the resonant cavity and are computationally economical. In addition to commercially available softwares such as COMSOL Multiphysics \cite{Shivkumar_2016} and ANSYS Fluent \cite{YAMADA_2006,Mesbahi_2013}, researchers all over the world also develop their own internal codes for optimization control, redundancy reduction, faster calculations for plasma chemistry by implementing various parallel algorithms and scalability \cite{Collin_thesis_2013,TAN_1995,Hasouni_1999,Mankelevich_2008,Hassouni_2010,Prasanna_2016,Prasanna_2017,ASHFOLD_2023}.         

 In the present work, we have used a commercially available software package (COMSOL Multiphysics) to design the reactor cavity and performed self-consistent steady-state simulations of the $H_{2}$ plasma excited using microwaves inside the cavity. Following the standard MPCVD reactor designing methodology mentioned in Ref. \cite{Silva_Hasouni_2009}, initially, we have selected the resonance mode (transverse Magnetic $TM_{011}$), coaxial microwave coupling structure, geometry of the quartz window. Next, we optimize the cavity dimensions by performing pure electromagnetic simulations, which solves Maxwell's equation in the absence of plasma ignition to get a resonance around 2.45 GHz. Cavity design is tweaked in such a way that we get a high electric field amplitude zone in front of the substrate stage. Afterwards, using plasma module of COMSOL package, we have obtained solutions for the hydrogen $H_{2}$ plasma discharge chemistry by solving mass, momentum and energy conservation equations i.e continuity, Navier-Stoke's and heat conduction equations respectively, along with various types of electron-$H_{2}$ chemical reactions which reflects composition and shape of the plasma inside the cavity. Once the reactor cavity design is optimized, we had proceeded for the fabrication process. 

 In addition, we have computationally investigated the effect of microwave power, gas pressure, and Coupled Effect of Microwave Power and Pressure variation on Plasma Characteristics such as electron density, gas temperature and atomic hydrogen density inside the newly designed cavity. It made convenient for us to realize the parametric space  for the reactor operations.  Also, validation of the design and simulation results were done by performing preliminary experiments which include vacuum integrity and hydrogen plasma ignition tests in the novel cavity. During these experiments, hemispherical stable plasma ball of pink in colour is observed which indicates that the fabricated reactor cavity can be used for diamond deposition.  
 
The paper is organized as follows: In Sec. \ref{simulation_methodology}, we present mathematical model equations implemented for the cavity designing. In the following subsections, i.e, Sec. \ref{EF_Module}, \ref{Plasma_Module} and \ref{Flow_thermal_Module}, we describe EM, plasma and flow-thermal modules respectively. Reactor design, prototype assembly along with optimization methodology is presented in Sec. \ref{new_reactor_design_prototype_assembly}. In Sec. \ref{simulation_result}, we present simulation results related to electromagnetic field, variation in power, pressure, and simultaneous power and pressure on plasma parameters inside the MPCVD reactor cavity. The results related to experimental validation and initial $H_{2}$ plasma ignition is shown in Sec. \ref{experimental_validation}. Finally, we conclude in Sec. \ref{Discussion and conclusion}. 

\section{Mathematical model and simulation methodology}
\label{simulation_methodology}

\begin{figure}
\centerline{\includegraphics[scale=0.30]{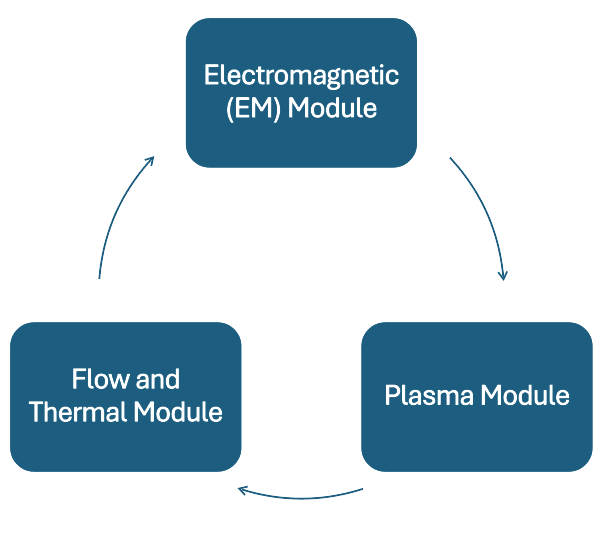}}
\caption{Portrait illustrating the multiphysics coupling between different modules used for whole device MPCVD reactor modeling. It consist of electromagnetic (EM), plasma, flow and thermal modules respectively.}
\label{FIG_1_Multiphysics_Coupling}
\end{figure}

As discussed earlier, the whole device modeling of MPCVD reactor cavity involves multi-physics simulation approach which consist of three major interconnected modules, namely, microwave electric field module, plasma module, flow and thermal module as shown in Fig. \ref{FIG_1_Multiphysics_Coupling}. Three different modules radio frequency (RF) module, plasma module, and heat transfer module of the commercial COMSOL Multiphysics software package have been used to perform the numerical simulations. In the following subsections, we will briefly describe about these individual modules and their coupling structure.

\subsection{Microwave electric field simulation module}
\label{EF_Module}
In a MPCVD reactor, Maxwell's equation is used to model the microwave electric field which is given as \cite{SU_2014,Shivkumar_2016,Wang_2019},
\begin{equation}
\label{EQ_1}
\bf
    \nabla \times \left[ \frac{1}{\mu_{r}} (\nabla \times  E) \right] = k_{0}^{2} \left[ \epsilon_{r} - \frac{j \sigma}{\omega \epsilon_{0}} \right] E 
\end{equation}
\begin{equation}
\label{EQ_2}
    k_{0}^{2} = \omega^{2} \epsilon_{0} \mu_{0} ~~~ with ~~~ \omega = 2\pi \times 2.45 \times 10^{9} ~s^{-1} 
\end{equation}
where $ k_{0} = 2\pi/\lambda_{0}$ is the wavenumber of microwave in free space, $\lambda_{0}$ is the microwave wavelength with a frequency of 2.45 GHz, $\omega$ is the microwave angular frequency, $\epsilon_{0}$ is the electric permittivity, $\sigma$ is the plasma conductivity, $\epsilon_{r}$ is the relative permittivity, $\mu_{r}$ is the relative permeability and $\bf E$ is the microwave electric field. In the presence and absence of the plasma, relative permittivity $\epsilon_{r}$ and plasma conductivity $\sigma$ should be determined as follows,\\

1. Regions where no plasma exists :
\begin{equation}
\label{EQ_3}
    \epsilon_{r}=1 ~~~~ and ~~~~ \sigma =0
\end{equation}

2. Regions where plasma exists \cite{FUNER_1995,Cherrington} :
\begin{equation}
\label{EQ_4}
    \epsilon_{r}=1 - \left[ \frac{\omega_{p}^{2}}{\omega^{2}+\nu_{e}^{2}} \right] ~~;~~ \sigma = \nu_{e} \left[ \frac{\epsilon_{0} \omega_{p}^{2}}{\omega^{2}+\nu_{e}^{2}} \right]
\end{equation}
Here, $\omega_{p}$ is the electron plasma frequency given as, $\omega_{p}^{2}= e^{2} n_{e}/ m_{e} \epsilon_{0}$ \cite{FUNER_1999}, $e$ and $m_{e}$ are the charge and mass of the electron, $n_{e}$ is the electron density, and $\nu_{e}$ is the electron collision frequency which can be determined using the following relation,
\begin{equation}
\label{EQ_5}
   \nu_{e} \approx a \left( \frac{p}{T_{g}} \right) ~~:~~ a= 1 \times 10^{10} K.Pa^{-1}.s^{-1}
\end{equation}
where $p$ and $T_{g}$ are the gas pressure and gas temperature of the hydrogen plasma respectively, and $a$ is the proportionality constant \cite{FUNER_1999}. 

\begin{table*}
\caption{Table for $H_{2}$ plasma discharge reactions involving electron interactions alongwith their energy estimates. In addition, plasma surface reactions with respective sticking coefficients used in the discharge simulations are also mentioned.}  
\centering                         
\begin{tabular}{c c c c c c}           
\hline\hline                        
 & Gas Phase Reactions &  &  Surface Reactions  &  \\ [1.0ex]
 \hline
Type & Reactions  & Energy (eV) & Reactions  & Sticking coefficients \\ [1.0ex] 
\hline          
Elastic & e$^{-}$ + H$_{2}$ $\longrightarrow$ e$^{-}$ + H$_{2}$  &  -  & H $\longrightarrow$ 0.5H$_{2}$ & 0.02  \\          
Dissociation & e$^{-}$ + H$_{2}$ $\longrightarrow$ e$^{-}$ + 2H & 7.923 - 11.72 & H(n=2) $\longrightarrow$ H & 1.0  \\ 
Excitation & e$^{-}$ + H$_{2}$ $\longrightarrow$ e$^{-}$ + H$_{2}$ & 12.4 - 14.6 & H(n=3) $\longrightarrow$ H & 1.0  \\
Ionization & e$^{-}$ + H$_{2}$ $\longrightarrow$ 2e$^{-}$ + H$_{2}^{+}$ & 15.4 & H$^{+}$ $\longrightarrow$ H & 1.0  \\
Elastic & e$^{-}$ + H $\longrightarrow$ e$^{-}$ + H & - & H$_{2}^{+}$ $\longrightarrow$ H$_{2}$ & 1.0  \\
Excitation & e$^{-}$ + H $\longrightarrow$ e$^{-}$ + H(n=2) & 10.20 & H$_{3}^{+}$ $\longrightarrow$ H$_{2}$ + H & 1.0  \\
Ionization & e$^{-}$ + H $\longrightarrow$ 2e$^{-}$ + H$^{+}$ & 13.60 &  &  & \\
Recombination & e$^{-}$ + H$_{2}^{+}$ $\longrightarrow$ H + H(N=3) & 0.01 &  &  & \\ 
Recombination & e$^{-}$ + H$_{3}^{+}$ $\longrightarrow$ 3H & 0.0 &  &  & \\
Recombination & e$^{-}$ + H$^{+}$ $\longrightarrow$ H(n=2) & 0.0 &  &  & \\
Recombination & e$^{-}$ + H$^{+}$ $\longrightarrow$ H(n=3) & 0.0 &  &  & \\
Ionization & H(n=2) + H$_{2}$ $\longrightarrow$ e$^{-}$ + H$_{3}^{+}$ & - &  &  & \\
Ionization & H(n=3) + H$_{2}$ $\longrightarrow$ e$^{-}$ + H$_{3}^{+}$ & - &  &  & \\
Ionization & H$_{2}$ + H$_{2}^{+}$ $\longrightarrow$ H + H$_{3}^{+}$ & - &  &  & \\ [1ex]
\hline\hline                               
\end{tabular}
\label{TABLE_1A}
\end{table*}

\subsection{Plasma simulation module}
\label{Plasma_Module}
Moderate pressure $H_{2}$ plasma discharges suitable for diamond deposition show strong chemical and thermal non-equilibrium characteristics. Hence, to simulate these type of plasmas is quite complex compared to the microwave electric field in the previous module. Ideally, one can perform computational modeling of such plasma discharge behavior by solving continuity, momentum transfer and energy balance equations for each chemical species generated in the discharge \cite{Hassouni_2010}. It also requires various input parameters for associated electron-electron and electron-heavy species interactions such as, species transport coefficients, reaction rate constant calculated from electron energy distribution function (EEDF) \cite{Hagelaar_2005}. One computes these parameters using various available Boltzmann solvers such as BOLSIG+ \cite{Hagelaar_2005}, ThunderBoltz \cite{Park_2024}, METHES \cite{RABIE_2016}, Magboltz \cite{BIAGI_1999}, LoKI–MC \cite{DIAS_2023}, BetaBoltz \cite{RENDA_2021} and LXCat datasets \cite{LC_Pitchford_2017,Carbone_2021}. However, several approximated and reduced models are used to minimize the computational rigor while modeling these low-temperature plasma discharges. Some of those approaches are briefly discussed.   

In previous works, Authors \cite{Shivkumar_2016,LI_2011} implemented Funer's model to estimate electron number density $n_{e}$. According to Funer's model electron number density is directly proportional to the generated local electric field inside the cavity given as \cite{YAMADA_2006,FUNER_1999,FUNER_1995},
\begin{equation}
\label{EQ_6}
\bf
   n_{e} = \left\{ \begin{array}{rcl}
             \gamma (|E - E_{min} |)+ n_{e,min} & \mbox{for} & E > E_{min} \\
             0 &  & otherwise
            \end{array}\right.
\end{equation}
where $\bf E_{min}$ is the minimum electric field needed to sustain a plasma discharge which is taken to be half of the breakdown voltage. In another attempt, Author \cite{SU_2014} applied a simple phenomenological model based on the drift-diffusion framework which assumes that the approximated plasma density at each point in the computational domain is a result of equilibrium between various electron generation source and loss sink terms given as,
\begin{equation}
\label{EQ_7}
\bf
   \nabla . (-D_{e} \nabla n_{e}) + R_{\nu r}.n_{e}^{2} + R_{a}.n_{e} = R_{i}.E^{2}.n_{e}
\end{equation}
where $ D_{e}$ is electron ambipolar diffusion coefficient, $\bf R_{i}$ is ionization coefficient of gas molecules due to electron collisions, $\bf R_{\nu r}$ is electron recombination coefficient, $\bf R_{a}$ is electrons to neutral particle attachment coefficient and $\bf n_{e}$ is the electron density\cite{SU_2014}.


In our present work, we have used Drift-Diffusion transport theory to simulate the $H_{2}$ plasma discharge by solving the following density and energy continuity equations given as,
\begin{equation}
\label{EQ_8}
\bf
\frac{\partial}{\partial t} (n_{e}) + \nabla . [-n_{e}(\mu_{e}.E)-D_{e}. \nabla n_{e} ] = R_{e}
\end{equation}
\begin{equation}
\label{EQ_9}
\bf
\frac{\partial}{\partial t} (n_{\epsilon}) + \nabla . [-n_{\epsilon}(\mu_{\epsilon}.E)-D_{\epsilon}. \nabla n_{\epsilon} ] + E . \Gamma_{\epsilon} = R_{\epsilon}
\end{equation}
where $\bf R_{e}$ and $\bf R_{\epsilon}$ are the electron source and energy loss terms due to inelastic collision, $\bf D_{e}$ and $\bf D_{\epsilon}$ are the electron and energy diffusivity constants, $\bf \mu_{e}$ nad $\bf \mu_{\epsilon}$ are the electron energy and mobility constants, $\bf n_{e}$ and $\bf n_{\epsilon}$ are the electron number and energy densities, $\bf \Gamma_{\epsilon}$ is the electron energy flux i.e $\bf { \Gamma_{\epsilon} = - (\mu_{\epsilon} . E ) n_{\epsilon} - \nabla (D_{\epsilon} n_{\epsilon}) }$ and $\bf E$ is the microwave electric field. Electron source term $\bf R_{e}$ is given as,
\begin{equation}
\label{EQ_10}
\bf
R_{e}= \sum_{j=1}^{M} \chi_{j} k_{j} N_{n} n_{e}
\end{equation}
where $M$ is number of reactions which contribute to the growth or decay of electron density, $\chi_{j}$ is mole fraction of the chemical species, $\bf N_{n}$ is total neutral number density, $\bf n_{e}$ is electron number density and $\bf k_{j}$ is the reaction rate coefficient for reaction j given as,
\begin{equation}
\label{EQ_11}
\bf 
     k_{j}= AT_{e}^{(n_{T})} exp \left[ \frac{-E_{a}}{T_{e}} \right]
\end{equation}
Here, $\bf A$ is pre-exponential factor which, in terms of the collision theory, is the frequency of correctly oriented collisions between the reacting species, $\bf T_{e}$ is electron temperature, $\bf n_{T}$ is electron temperature exponent, $\bf E_{a}$ is the activation energy (in eV). In Table \ref{TABLE_1A}, we have listed various $H_{2}$ plasma discharge reactions associated with elastic, ionization, excitation, dissociation, recombination chemical reactions. Also, reactions to model plasma surface interactions with their respective sticking coefficients are also tabulated. 

\subsection{Flow and Thermal module}
\label{Flow_thermal_Module}

In order to investigate the fluid flow dynamics inside the MPCVD reactor, we numerically solve mass and momentum conservation equations i.e continuity and Navier-Stokes equations, respectively, given as,
\begin{equation}
\label{EQ_12}
\bf 
\frac{\partial \rho}{\partial t} + \nabla . (\rho u) = 0
\end{equation}
\begin{equation}
\label{EQ_13}
\bf 
\rho \frac{\partial u}{\partial t} + \rho (u.\nabla)u = \nabla.[ -p I + \tau] + F
\end{equation}
where $\rho$ is density, $u$ is velocity field vector, $p$ is pressure, $I$ is identity matrix, $\bf \tau$ is the viscous stress tensor and $\bf F$ is the volumetric force vector. Under the 2D axisymmetric formulation i.e absence of any gradients in the azimuthal direction $\phi$ and time independent steady state assumptions,
\begin{equation}
\label{EQ_14}
\partial / \partial \phi \longrightarrow 0 ~~;~~ u_{\phi}=0 ~~;~~ \partial / \partial t \longrightarrow 0
\end{equation}
Eqs. \ref{EQ_12} and \ref{EQ_13} result into,
\begin{equation}
\label{EQ_15}
\bf
\nabla . (\rho u) = 0
\end{equation}
\begin{equation}
\label{EQ_16}
\bf
\rho (u.\nabla)u = \nabla.[ -p I + \tau] + F
\end{equation}
Hence, Eqs. \ref{EQ_15} and \ref{EQ_16} are the reduced Navier-Stokes equation used to investigate the laminar fluid flow dynamics in the cylindrical cavity MPCVD reactor. There are a couple of key advantages of the resulting reduced system of equations, i.e, it is less computationally expensive and has better convergence of the solutions compared to retaining the equations in $\phi$-direction. 

At low pressures, gas convection in the vacuum chamber could be ignored as diffusion transport is relatively fast, thermal distributions inside the cavity can be sampled using heat conduction or conservation of energy equation given as,
\begin{equation}
\label{EQ_17}
\bf
\rho C_{p} \left[ \frac{\partial T}{\partial t} + (u.\nabla) T \right]   =  - (\nabla . q) + \tau : S -  \\ \left.\frac{T}{\rho} \frac{\partial \rho}{ \partial T}\right\vert_{p}  \left( \frac{\partial p}{\partial t} + (u. \nabla)p  \right) + Q
\end{equation}
where $C_{p}$ is the specific heat capacity at constant pressure, $\rho$ is density, $\bf u$ is the velocity vector, $T$ is absolute temperature, $\bf q$ is the heat flux vector, $Q$ is the heat source, $\bf \tau$ is the viscous stress tensor, $\bf S$ is the strain-rate tensor defined as,
\begin{equation}
\label{EQ_18}
\bf 
    S = \frac{1}{2} \left[ \nabla u + (\nabla u)^{T} \right]
\end{equation}
The operation ``:" denotes a contraction between tensors or referred to as the double dot product defined by,
\begin{equation}
\label{EQ_19}
    p : q = \sum_{n} \sum_{m} p_{nm} q_{nm}
\end{equation}

Eqs. \ref{EQ_1},\ref{EQ_8},\ref{EQ_9},\ref{EQ_15},\ref{EQ_16} and \ref{EQ_17} are solved using commercially available software COMSOL Multiphysics which is based on finite element method (FEM) fluid simulation approach. We excite transverse magnetic ($TM_{01}$) mode inside the resonant cavity by imposing zero tangential component of the electric field boundary conditions on the cavity walls ensuring normal incidence of the field on the surface. To discritize the computational domain, physics based controlled meshing is used with greater refinement in the corner regions of the geometry. In the next section, we present our novel cylindrical MPCVD cavity design and discuss the key parameters that plays an important role in the MPCVD reactor design process.   

\section{New reactor design and prototype assembly}
\label{new_reactor_design_prototype_assembly}

\begin{figure*}
\centerline{\includegraphics[scale=0.40]{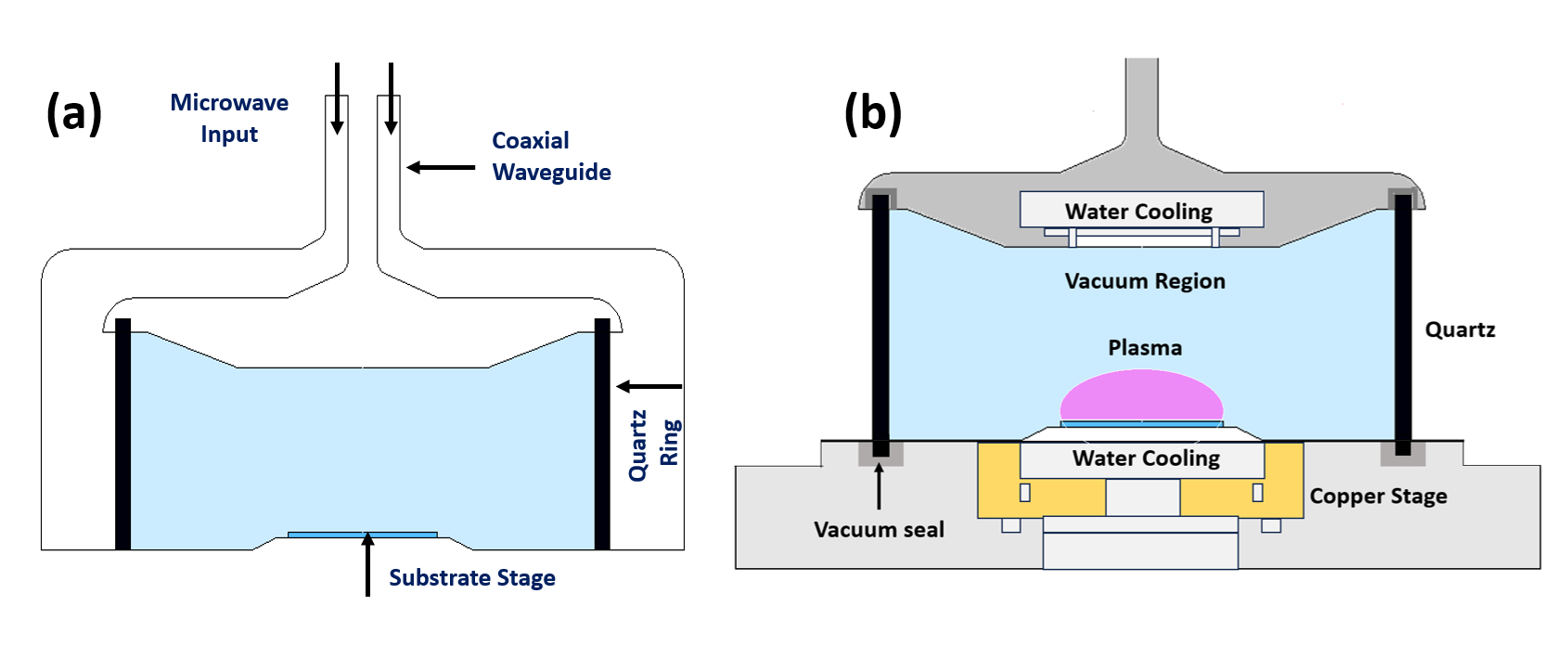}}
\caption{Simulation domain used in COMSOL multiphysics software and cross-sectional schematic of the newly designed cylindrical MPCVD reactor vacuum cavity using quartz ring. Figs. (a) and (b) indicate the arrangement of critical components in the reactor such as top launch microwave power input, coaxial waveguide, copper substrate stage, quartz ring and thermal cooling area etc.  }
\label{FIG_2_Simulation_Domain_and_Schematic}
\end{figure*}

\begin{figure*}
\centerline{\includegraphics[scale=0.35]{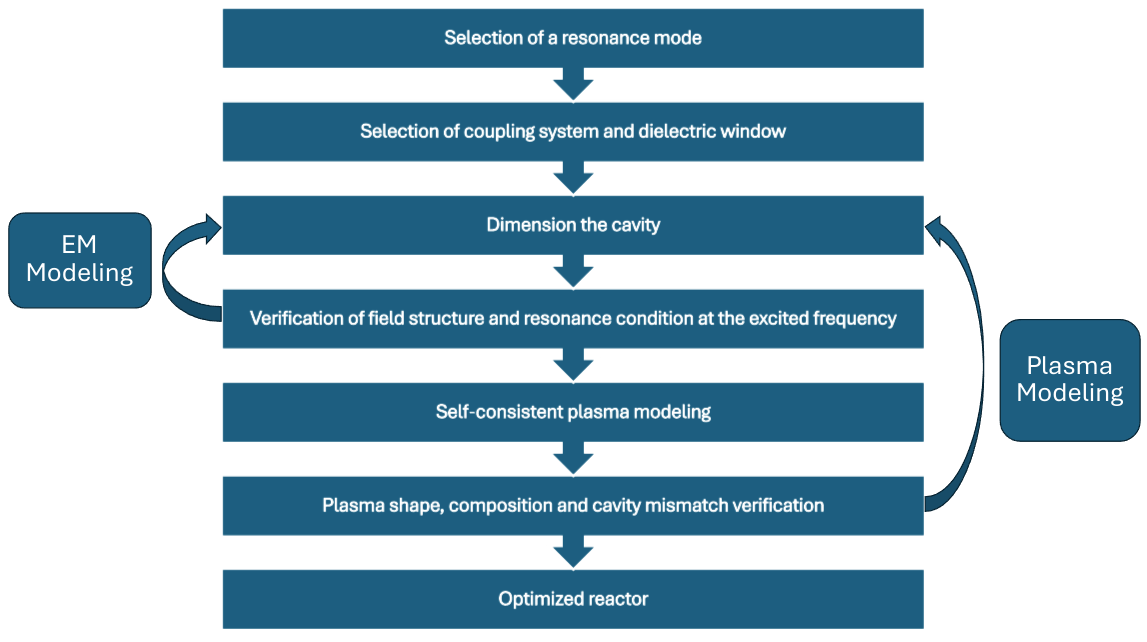}}
\caption{Flow chart showing optimization and design methodology for cylindrical MPCVD reactor, relying on the coupling between electromagnetic modeling and self-consistent plasma modeling.}
\label{FIG_3_reactor_design_flowchart}
\end{figure*}

 MPCVD cavities are basically resonant electromagnetic cavities with appropriate field structures suitable for diamond deposition. Plasma-assisted CVD process for diamond deposition requires relatively high pressures compared to the other material deposition techniques. Since, diffusion coefficient is inversely proportional to the pressure, diffusion phenomenon inside these cavities are limited, so, producing plasma and the associated high electric field region near the deposition substrate is a necessity. Couple of advantages of these resonant electromagnetic cavities are mentioned below,
\begin{itemize}
    \item Due to resonance phenomenon, if one excites an electromagnetic cavity at its resonance frequency (which depends on the dielectric constant and cavity geometry), it holds higher local electric field facilitating easier plasma ignition.   

    \item One can approximately choose to promote a high electric field structure in the region where plasma is to be ignited without using any electrodes. Hence, reducing the electrode contamination. 
\end{itemize}

The most important aspect that one has to take care while designing the microwave plasma reactor is to maintain the electric field structure such that it enhances the local electric field strength over the substrate and obtain largest possible region for the diamond deposition. Also, ignition of the plasma perturbs the cavity tuning which has to be taken into account from the initial designing stages.  

Depending on the orthogonality of the electric or magnetic field vectors with respect to the cavity axis, one can excite two different types of resonant modes, namely transverse electric (TE) or transverse magnetic (TM) modes. The main reason for choosing a TM mode for the MPCVD reactor cavity design is that, unlike TE modes, TM modes can provide strong axial electric fields that extend toward the substrate holder. TE modes have electric fields that are primarily transverse and vanish at the metallic walls, which limits their ability to generate high field intensities near the substrate surface, making plasma ignition and maintenance near the substrate less efficient . Generally, in most of the MPCVD reactors, microwave energy is transmitted with TE mode using rectangular waveguide and is converted to TM mode in the cavity through co-axial mode converter. From purely electromagnetic point of view, following are the crucial steps to design a microwave plasma reactor,
\begin{itemize}
    \item Resonant mode with appropriate cavity structure.
    \item Selection of coupling structure.
    \item Shape and location of the dielectric window.
\end{itemize}

\begin{figure*}
\centerline{\includegraphics[scale=0.70]{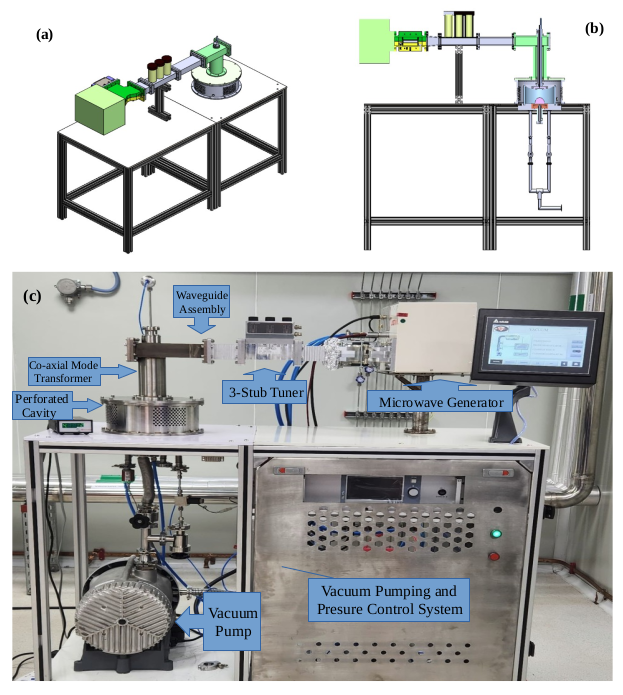}}
\caption{The new cylindrical MPCVD reactor (a) top view, (b) side view and (c) actual image of the reactor prototype assembly. In Fig. (c) microwave generator, microwave passive component assembly i.e waveguide assembly and 3-stub tuner, co-axial mode converter, perforated vacuum cavity, vacuum pump along with pressure control systems are indicated.}
\label{FIG_4_Prototype_Assembly}
\end{figure*}

In the present reactor design, we have taken into account of all the above mentioned factors associated with microwave plasma reactor design. Fig. \ref{FIG_3_reactor_design_flowchart} shows the complete optimization and design methodology consisting of electromagnetic and plasma modeling implemented for the reactor design. At first, we have chosen the resonance mode, microwave coupling structure, geometry and location of the dielectric window. Secondly, we optimize the cavity dimensions using electromagnetic solver in order to get a cavity resonance at 2.45 GHz along with ensuring a large electric field zone in front of the substrate holder. Afterwards, solutions for the hydrogen $H_{2}$ plasma discharge is carried out inside the cavity, with the help of self-consistent plasma module as discussed in Sec. \ref{Plasma_Module}, resulting into composition and shape of the plasma, as well as cavity detuning. The complete process is iterated by modifying the cavity dimensions, if the results are not satisfactory. This optimization is made under high microwave power deposition (MWPD), under high pressure discharge conditions, as one can easily perform corrections for the detuning at low power density. One can clearly notice the heavy reliance of the reactor design methodology on electromagnetic and self-consistent plasma modeling, hence, it becomes absolutely mandatory to ensure that the EM and plasma modules yield converged solutions within required tolerance during numerical simulations.

Fig. \ref{FIG_2_Simulation_Domain_and_Schematic} (b) illustrates the cross-sectional schematic diagram of the novel cylindrical cavity consisting of microwave antenna, cooling management area, quartz ring for vacuum seal, copper stage and Fig. \ref{FIG_4_Prototype_Assembly} illustrates the (a) top view, (b) side view and (c) image of the actual reactor prototype assembly including microwave power source, waveguide arrangement, perforated vacuum chamber and vacuum pumping and pressure control systems. The developed reactor is a top-launch system composed of perforated stainless-steel cylindrical vacuum chamber with a volume of approximately 3.0 L. To make the distribution of microwave power more uniform, the chosen cavity is entirely axi-symmetrical. The co-axial mode transformer or waveguide, located on the top of the cavity, was designed as a unique antenna. Cylindrical quartz is used to affix the antenna as well as provide the vacuum seal for the cavity. The sealing reliability of the cavity is provided by pressure difference between inside and outside of the cavity during the operation. Also, gravity of the antenna could enhance the sealing of the cavity. Gas inlet inside the cavity is facilitated by the antenna itself. Its design consists of two aspects,
\begin{itemize}
    \item Co-axial tube that allows the gas to flow through the chamber.
    \item Co-axial tube that allows the air to flow through the antenna in order to keep the antenna in the atmospheric pressure which also allows heat transfer from the antenna.
\end{itemize}
Generally, in most of the MPCVD reactors, thermal management is the main concern for the long time and high power operations. We have designed both our unique antenna and copper stage in such a way that sufficent area is provided for proper cooling arrangement as indicated in Fig. \ref{FIG_2_Simulation_Domain_and_Schematic} (b). Rest of the reactor dimensions are given below,
\begin{itemize}
    \item Inner diameter of the quartz cavity - 180 mm.
    \item Outer diameter of the quartz cavity - 192 mm.
    \item Height of the quartz cavity - 100 mm.
    \item Inner diameter of the perforated cavity - 265 mm.
    \item Diameter of Substrate Holder - 100 mm.
    \item Height of the Substrate Holder - 18 mm.
    \item Diameter of the perforation holes - 5 mm.
    \item Diameter of the antenna - 205 mm.
    \item Diameter of the coaxial tube - 24 mm.
    \item Diameter of the Molybdenum substrate - 52.8 mm.
    \item Height of the Molybdenum substrate - 3 mm.
\end{itemize}
In the next section, we show the results of parametric simulation investigations which led to the optimization of the current MPCVD reactor design discussed in this section.

\section{Simulation Results}
\label{simulation_result}

In this Section, we present the simulation results which include variation of electromagnetic (EM) field, individual effect of power and pressure variation, and coupled effect of power power-pressure variations on the $H_{2}$ plasma discharge parameters in the newly designed perforated microwave cavity.

\subsection{Electromagnetic (EM) Simulations: In the absence of $H_{2}$ Plasma Ignition}
\label{electromagnetic_simulation_results}

\begin{figure*}
\centerline{\includegraphics[scale=0.35]{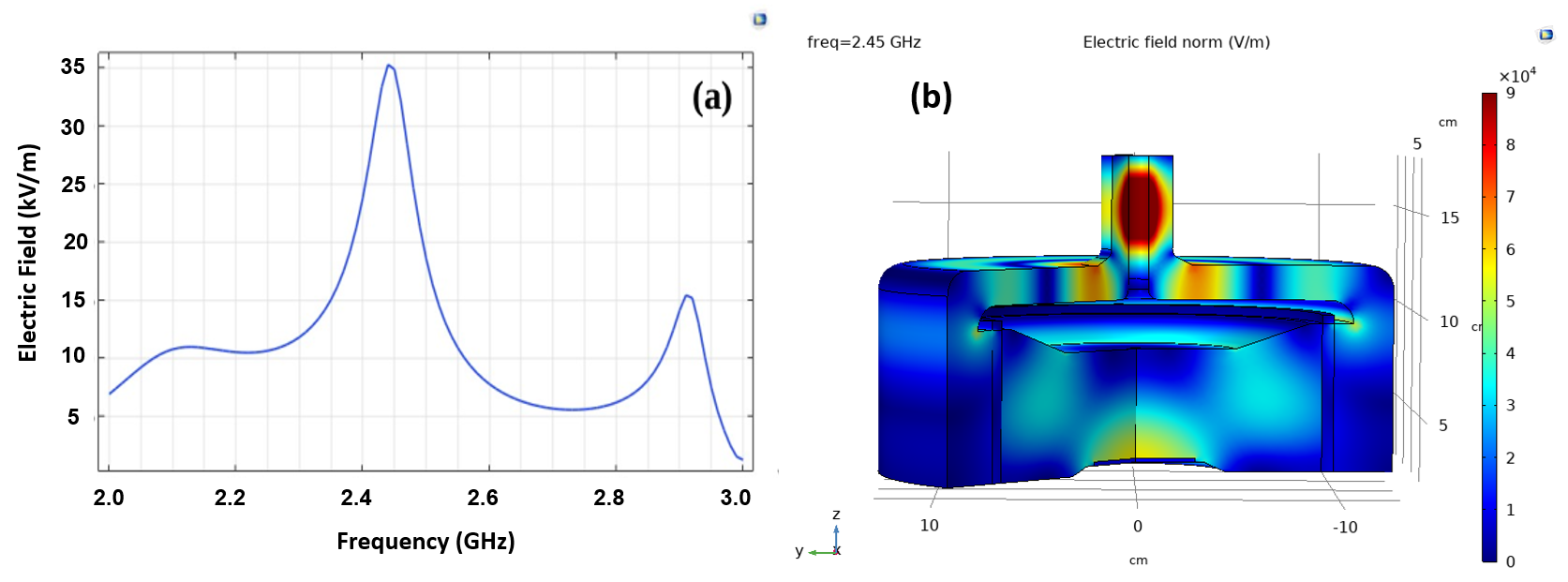}}
\caption{1D plot for average electric field intensity Vs the microwave frequency over the substrate stage, and (b)
3D E-Field contour inside the cavity resonating at 2.45 GHz Frequency in vacuum conditions i.e. in the absence of $H_{2}$ plasma ignition.}
\label{FIG_5_Vacuum_electric_field}
\end{figure*}


\begin{figure*}
\centerline{\includegraphics[scale=0.35]{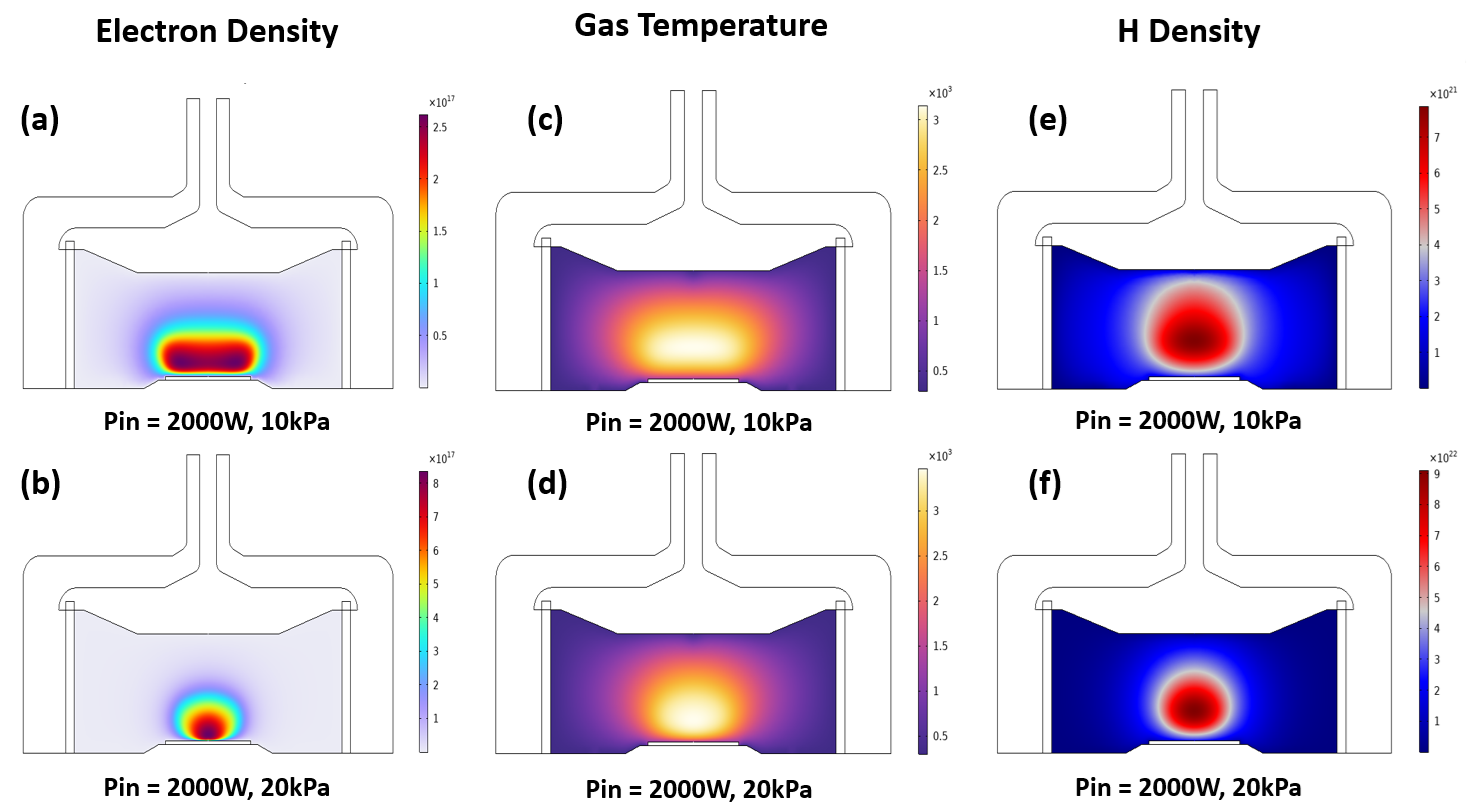}}
\caption{2D (r,z) contour signature of the hydrogen $H_{2}$ plasma discharge parameters at constant applied microwave power with various pressure values i.e, (a) electron density at $P_{in}=2$ kW, $p_{0}=10$ kPa, (b) electron density at $P_{in}=2$ kW, $p_{0}=20$ kPa, (c) gas temperature at $P_{in}=2$ kW, $p_{0}=10$ kPa, (d) gas temperature at $P_{in}=2$ kW, $p_{0}=20$ kPa, (e) hydrogen $H_{2}$ number density at $P_{in}=2$ kW, $p_{0}=10$ kPa and (f) hydrogen $H_{2}$ number density at $P_{in}=2$ kW, $p_{0}=20$ kPa.}
\label{FIG_7_2D_PLOT_POWER_CONST_PRESURE_VARY}
\end{figure*}

To begin with, we followed the reactor design and optimization methodology mentioned in Sec. \ref{new_reactor_design_prototype_assembly}. Firstly, we optimize the reactor geometry by modeling the distribution of microwave electric field inside the reactor using COMSOL Multiphysics software package, which solves the governing Maxwell's equations i.e., Eqs. \ref{EQ_1} and \ref{EQ_2} discussed in \ref{EF_Module} by finite element method (FEM). Using the pure electromagnetic (EM) simulations i.e., in the absence of the $H_{2}$ plasma discharge, we excite transverse magnetic $TM_{01}$ mode in the cylindrical cavity and finalized the optimized reactor geometry whose schematic is shown in Fig. \ref{FIG_2_Simulation_Domain_and_Schematic}, Fig. \ref{FIG_4_Prototype_Assembly} and dimension is given in Sec. \ref{new_reactor_design_prototype_assembly}.
 
 Fig. \ref{FIG_5_Vacuum_electric_field} shows (a) 1D plot for Electric field intensity Vs the microwave frequency inside the cavity, and (b) 3D E-Field contour inside the cavity resonating at 2.45 GHz Frequency. Fig. \ref{FIG_5_Vacuum_electric_field} (a) indicates that the maximum electric field amplitude achieved corresponds to 2.45 GHz of the applied microwave frequency i.e. the cavity structure resonates quite well on the desired microwave frequency. From Fig. \ref{FIG_5_Vacuum_electric_field} (b), we inferred that inside the vacuum region the major fraction of the microwave energy coupled over the above the substrate copper stage only, which is a favorable configuration for the diamond deposition process. Also, as per our simulation observation, the maximum amplitude of the microwave electric field intensity inside the cavity above the copper stage is $3.5 \times 10^{4}$ V/m. 

 Consequently, by fetching the required output from the EM module as an input to the plasma module, we perform self-consistent plasma modeling by solving mass, momentum and energy conservation equations described in plasma module (Sec. \ref{Plasma_Module}) and flow-thermal module (Sec. \ref{Flow_thermal_Module}) under the 2D axisymmetric formalism for hydrogen $H_{2}$ plasma discharge inside the cavity. For the present cavity design, our computational investigations are mainly aimed at tracing the effect of power, pressure, and coupled effect of power and pressure variations on plasma characteristics which is presented in the following subsections.

\subsection{Effect of pressure variation on the $H_{2}$ plasma characteristics}
\label{Effect_of_pressure_variation_on_plasma_characteristics}

In these parametric studies, we have performed the simulation for fixed applied microwave power at $P_{in}=2$ kW with two different pressure values i.e $p_{0}= 10 ,~ 20$ kPa. Fig. \ref{FIG_7_2D_PLOT_POWER_CONST_PRESURE_VARY} shows the 2D $(r,z)$ variation of hydrogen $H_{2}$ plasma discharge parameters at a applied microwave power of $P_{in}=2$ kW with two different pressure values i.e, (a) electron density at $p_{0}=10$ kPa, (b) electron density $p_{0}=20$ kPa, (c) gas temperature at $p_{0}=10$ kPa, (d) gas temperature at $p_{0}=20$ kPa, (e) hydrogen $H_{2}$ number density at $p_{0}=10$ kPa and (f) hydrogen $H_{2}$ number density at $p_{0}=20$ kPa. Generally, increase in the pressure of the gas inside the cavity leads to an increase in the collision rate (electron collide more often) which further increases the energy dissipation into the system. As a result, the density of electrons is increased, which reduces the plasma conductivity and increases the depth of penetration of MW, resulting in a decrease in the amplitude of the microwave electric field \cite{Shivkumar_2016}. Also, one observes the concentration of the plasma shrink into a smaller region above substrate stage due to increase in the pressure at a fixed applied microwave power.

\begin{figure*}
\centerline{\includegraphics[scale=0.36]{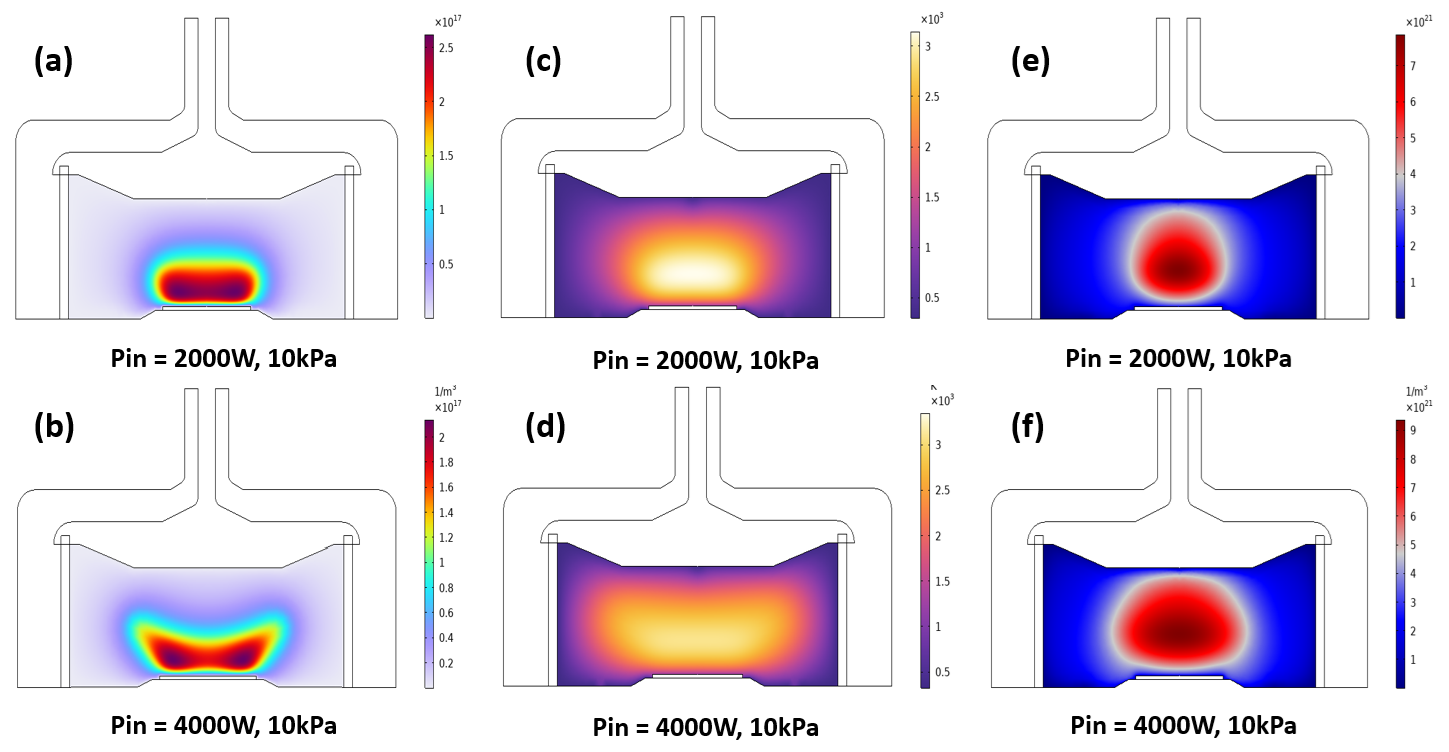}}
\caption{2D (r,z) contour variation of plasma parameters at constant pressure with various applied microwave power i.e, (a) electron density $n_{e}$ at $P_{in}=2$ kW, $p_{0}=10$ kPa, (b) electron density at $P_{in}=4$ kW, $p_{0}=10$ kPa, (c) gas temperature $T_{g}$ at $P_{in}=2$ kW, $p_{0}=10$ kPa, (d) gas temperature at $P_{in}=4$ kW, $p_{0}=10$ kPa, (e) hydrogen $H_{2}$ number density $n_{H}$ at $P_{in}=2$ kW, $p_{0}=10$ kPa and (f) hydrogen $H_{2}$ number density at $P_{in}=4$ kW, $p_{0}=10$ kPa.}
\label{FIG_8_2D_PLOT_PRESURE_CONST_POWER_VARY}
\end{figure*}

In Fig. \ref{FIG_7_2D_PLOT_POWER_CONST_PRESURE_VARY} (a) and (b), at a fixed power value $P_{in}=2$ kW, we have observed an increase in the 2D $(r,z)$ contour electron density $n_{e}$ signature inside the cavity with the increase of applied pressure $p_{0}$ from $p_{0}=10$ kPa to $p_{0}=20$ kPa. Also, increasing the pressure seems to concentrate the plasma at the center in a smaller region above the copper stage, thereby increasing the $n_{e}$ in the cavity. Symmetric variations in the 2D $n_{e}(r,z)$ signature can be seen along the axial $(z)$ directions within the plasma region. Electron temperature $T_{e}$ is directly proportional to the electric field intensity inside the cavity and inversely proportional to the applied gas pressure $p_{0}$ at a constant microwave power. On the other side, gas temperature $T_{g}$ has a very strong dependence on the applied $p_{0}$ inside the cavity region. An increment in the $p_{0}$ results into a reduction in the electron temperature but increases the number of collisions among the gas molecules which led to more energy transfer between them resulting into an increment in the gas temperature $T_{g}$. In Fig. \ref{FIG_7_2D_PLOT_POWER_CONST_PRESURE_VARY} (c) and (d), we observe that with the increase in the gas pressure from $p_{0}=10$ kPa to $p_{0}=20$ kPa at a fixed applied power of $P_{in}=2$ kW, gas temperature also increases particularly above the substrate area inside plasma region. Gas temperature variations seems quite symmetric along $(r,z)$-directions within the plasma region. In fact, shrinking of the plasma region to a smaller volume while increasing the gas pressure is a notable factor in the reduction of the overall average gas temperature inside the cavity. 

In Fig. \ref{FIG_7_2D_PLOT_POWER_CONST_PRESURE_VARY} (e) and (f) indicates an increase in the atomic hydrogen number density $(n_{H})$ with an increase of gas pressure from $p_{0}=10$ kPa to $p_{0}=20$ kPa at a constant applied microwave power of $P_{in}=2$ kW. We observe symmetric variations of the hydrogen number density $n_{H}$ along both the axial and radial directions within the plasma region on increasing the gas pressure $p_{0}$. In the next Section, we present the simulation results associated with effect of microwave power variations on the $H_{2}$ plasma characteristics at a constant gas pressure of $p_{0}= 10$ kPa.  
\subsection{Effect of power variation on the $H_{2}$ plasma characteristics}
\label{Effect_of_power_variation_on_plasma_characteristics}

In order to perform these parametric investigation, we have set the simulation parameters for fixed gas pressure $p_{0}=10$ kPa and increased the applied microwave power i.e, from $P_{in}=2$ kW to $P_{in}=4$ kW. Fig. \ref{FIG_8_2D_PLOT_PRESURE_CONST_POWER_VARY} demonstrates two dimensional $(r,z)$ contour variation of plasma parameters at constant pressure $p_{0}=10$ kPa with different applied microwave power i.e, (a) electron density $n_{e}$ at $P_{in}=2$ kW, (b) $n_{e}$ at $P_{in}=4$ kW, (c) gas temperature $T_{g}$ at $P_{in}=2$ kW, (d) $T_{g}$ at $P_{in}=4$ kW, (e) hydrogen number density $n_{H}$ at $P_{in}=2$ kW, and (f) $n_{H}$ at $P_{in}=4$ kW.

\begin{figure*}
\centerline{\includegraphics[scale=0.42]{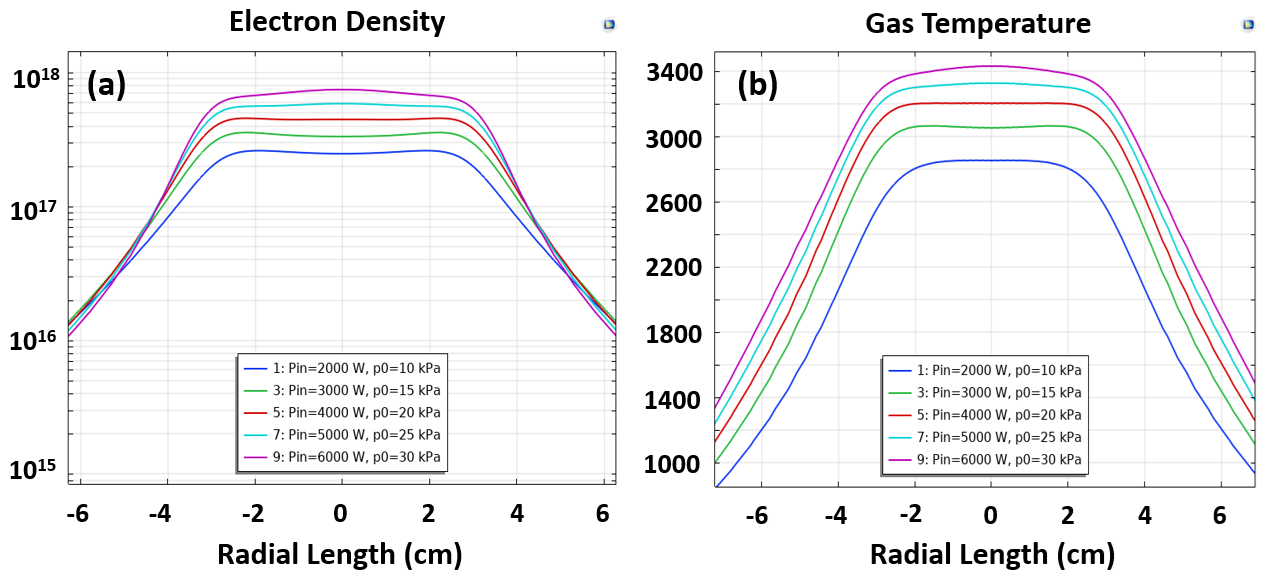}}
\caption{1D Variation of (a) electron density $n_{e}$, (b) gas temperature $T_{g}$ in plasma core along the radial $(r)$ direction for various power density i.e, power - pressure parameter combinations.} 
\label{FIG_9_1D_PLOT_Power_density_variations}
\end{figure*}



Unlike the effects observed with pressure variation, increasing the input MW power $P_{in}$ at a fixed gas pressure leads to an expansion of the plasma, accompanied by slight reductions in electron density.  In our COMSOL simulation results, similar effect can be seen in Fig. \ref{FIG_8_2D_PLOT_PRESURE_CONST_POWER_VARY} (a) and (b) respectively, where we have increased the microwave power from $P_{in}=2$ kW to $P_{in}=4$ kW at constant $p_{0}= 10~kPa$. We observed that the electron density ($n_{e}$) decreases slightly with increasing microwave power under constant pressure conditions in the MPCVD chamber. However, a radial non-uniformity in plasma density becomes more pronounced with the rise in power amplitude. Specifically, at higher power levels, the plasma density profile evolves from a centrally peaked distribution to one exhibiting two off-axis high-density regions—specifically towards the substrate edge. This bifurcation behavior has been attributed to changes in the electromagnetic field distribution within the plasma chamber, especially under conditions of higher plasma loading.At higher powers, the increased plasma conductivity can modify the local permittivity and absorb more power near the periphery, altering the standing wave or TM mode structure in the reactor. Therefore, the plasma density distribution exhibits significant radial non-uniformity. Notably, the peak density shifts away from the center, resulting in two off-axis high-density lobes.

\begin{figure*}
\centerline{\includegraphics[scale=0.45]{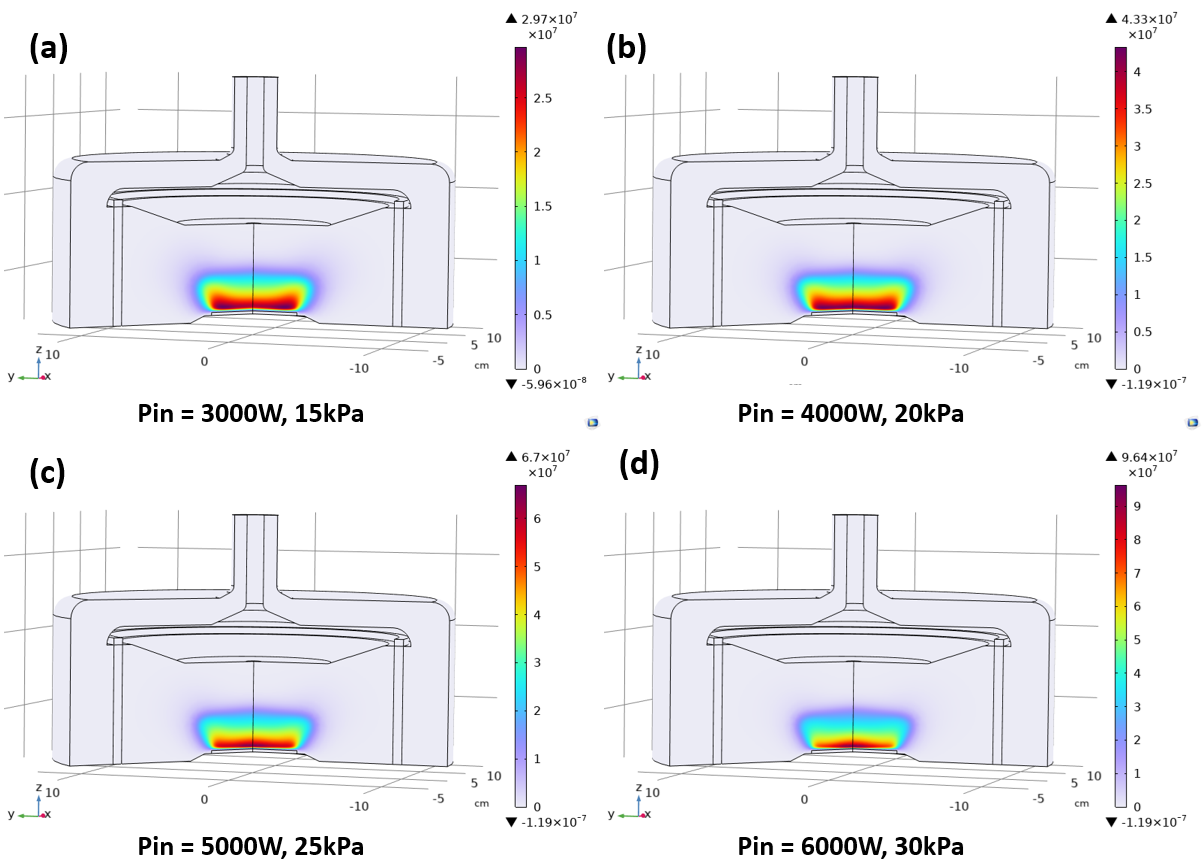}}
\caption{3D Contour variation of microwave power deposition (MWPD) above the substrate stage for various power and pressure parameters i.e (a)$P_{in}=3000$ W, $p_{0}=15$ kPa (b) $P_{in}=4000$ W, $p_{0}=20$ kPa (c) $P_{in}=5000$ W, $p_{0}=25$ kPa and (d) $P_{in}=6000$ W, $p_{0}=30$ kPa. It can be clearly seen that with the increase in the power density i.e pressure and power parameters, amplitude and area of the MWPD is increased around substrate stage.}
\label{FIG_10_MWPD_VARIOUS_POWER_PRESURE}
\end{figure*}

Electron temperature $T_{e}$ increases with an increment in the microwave power at constant gas pressure, as  more microwave energy is coupled into the plasma. This leads to stronger ionization and heating, causing the plasma core to expand. Since,  at higher powers, the microwave field sustains a larger plasma volume. The deposited energy is distributed over a broader region, so local heating at a single point such as plasma core becomes less intense. \cite{Shivkumar_2016}. In Fig. \ref{FIG_8_2D_PLOT_PRESURE_CONST_POWER_VARY} (c) and (d), we have observed slightly decrease in gas temperature in the plasma core as well as expansion in the 2D $(r,z)$ gas temperature contour signatures on increasing power value from $P_{in}=2$ kW to $P_{in}=4$ kW at constant $p_{0}$. Also, $T_{g}$ illustrates symmetric variations along both the $(r,z)$-directions with increase in the microwave power amplitude. Upon comparing the results of both Sections \ref{Effect_of_pressure_variation_on_plasma_characteristics} and \ref{Effect_of_power_variation_on_plasma_characteristics}, we can clearly conclude that $T_{g}$ has a week dependence on applied the microwave power and a strong dependence on the gas pressure as reported in Ref. \cite{Shivkumar_2016}. Fig. \ref{FIG_8_2D_PLOT_PRESURE_CONST_POWER_VARY} (e) and (f) illustrates an increment and enhancement in the atomic hydrogen number density $n_{H}$ with an increase of applied microwave power from $P_{in}=2$ kW to $P_{in}=4$ kW at constant $p_{0}= 10$ kPa. We have observed a symmetric enhancement of $n_{H}$ along both axial and radial directions within the plasma region above copper stage inside the cavity.

\subsection{Coupled Effect of Microwave Power and Pressure on Plasma Characteristics}
\label{Effect_of_power_density_variation_on_plasma_characteristics}

In the studies presented in the previous sections, applied power and pressure are varied by keeping other parameter constant and vice-versa. However, variation of these parameters in this way is not favorable for diamond deposition as we can observe the abrupt axial and radial change in the plasma shape, volume and density profiles above the substrate stage. Ideally, one should vary the power and pressure simultaneously, to focus the plasma formation uniformly above the substrate or deposition area. In one of the previous work, Authors \cite{Hassouni_2010} have illustrated that the electron number density variation while keeping the constant plasma volume which requires simultaneous change in the gas pressure and microwave power parameters. 

\begin{table}
\caption{Table for various applied microwave power [$P_{in}$] and gas pressure [$p_{0}$] values used for perforated cavity $H_{2}$ plasma optimization}  
\centering                         
\begin{tabular}{c c c}           
\hline\hline                        
Run No. & Applied power (W)  & Pressure (kPa)  \\    
        &   [$P_{in}$]       &   [$p_{0}$]  \\ [1.0ex] 
\hline          
Run 1 & 1000 & 10.0 \\          
Run 2 & 2000 & 10.0 \\
Run 3 & 2500 & 12.0 \\
Run 4 & 2500 & 13.5  \\
Run 5 & 3000 & 15.0 \\
Run 6 & 4000 & 20.0 \\
Run 7 & 4500 & 22.5 \\
Run 8 & 4500 & 23.0 \\
Run 9 & 5000 & 25.0 \\
Run 10 & 5500 & 27.5 \\
Run 11 & 5500 & 28.5 \\
Run 12 & 6000 & 30.0 \\ [1ex]
\hline\hline                               
\end{tabular}
\label{TABLE 1}
\end{table}

\begin{figure*}
\centerline{\includegraphics[scale=0.46]{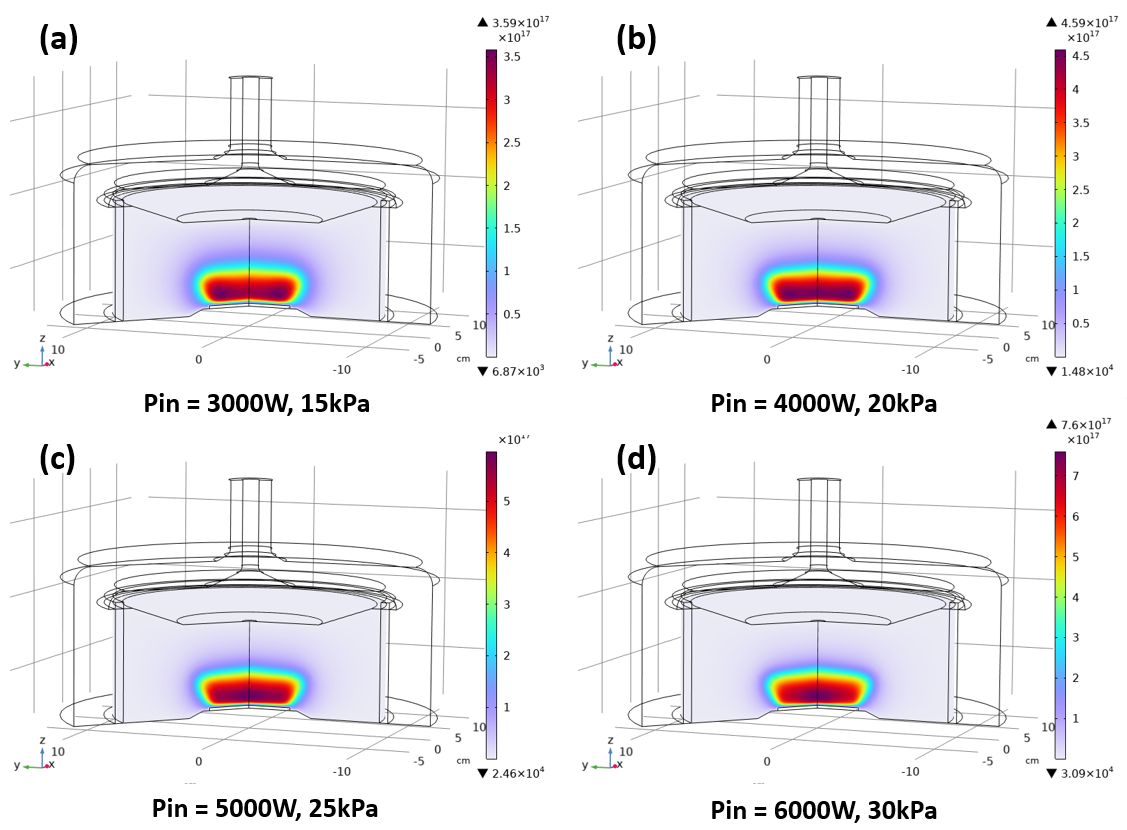}}
\caption{Portrait of 3D electron number density variations above the substrate stage for various power and pressure parameters i.e (a)$P_{in}=3000$ W, $p_{0}=15$ kPa (b) $P_{in}=4000$ W, $p_{0}=20$ kPa (c) $P_{in}=5000$ W, $p_{0}=25$ kPa and (d) $P_{in}=6000$ W, $p_{0}=30$ kPa. It can be clearly seen that with the increase in the power density i.e pressure and power parameters, electron number density is increased.}
\label{FIG_11_ELECTRON_DENSITY_VARIOUS_POWER_PRESURE}
\end{figure*}

In Table. \ref{TABLE 1}, we have tabulated the different applied microwave power [$P_{in}$] and gas pressure [$p_{0}$] parameter combinations used for $H_{2}$ plasma optimization in the newly designed perforated cavity. We have performed a total of 12 simulation run combinations where microwave power ranges from $P_{in}=1$ to $P_{in}=6$ kW and gas pressure ranges from $p_{0}=10$ to $p_{0}=30$ kPa. Fig. \ref{FIG_9_1D_PLOT_Power_density_variations} shows 1D Variation of (a) electron density $n_{e}$, (b) gas temperature $T_{g}$ in plasma core and with respect to radial $(r)$ length near substrate for various power density i.e, power-pressure combinations listed in Table. \ref{TABLE 1}. From Fig. \ref{FIG_9_1D_PLOT_Power_density_variations} (a), we have observed that the order of electron density $n_{e}$ ranges around $10^{17}~m^{-3}$ and with the simultaneous increase in the power pressure combinations it increases simultaneously. Also, $n_{e}$ density profile variations decreases sharply near the edge of the moly substrate i.e around 60 mm. The Fig. \ref{FIG_9_1D_PLOT_Power_density_variations} (b) indicate that along radial direction the gas temperature $T_{g}$ in the plasma core increases with increase in the power density parameters. The gas temperature in the plasma core reaches above 3400 K which is in agreement to the observations reported in Ref. \cite{Hemawan_2015}. Generally, the gas temperature $T_{g}$ near the substrate is thermally quenched by the water cooling arrangement of the substrate. In addition, we have seen the drop in the plasma core $T_{g}$ just after 55 mm while scanning radially outwards. Also, these simulation results helped us to understand that one can not easily attain uniform and higher amplitudes of $n_{e}$, $T_{g}$ profile until proper recipe i.e., proper combination of process parameter (power and pressure) is used.    

Fig. \ref{FIG_10_MWPD_VARIOUS_POWER_PRESURE} illustrates variation of microwave power deposition (MWPD) above the substrate stage inside the cavity for various power and pressure parameters tabulated in Table. \ref{TABLE 1} i.e (a)$P_{in}=3000$ W, $p_{0}=15$ kPa (b) $P_{in}=4000$ W, $p_{0}=20$ kPa (c) $P_{in}=5000$ W, $p_{0}=25$ kPa and (d) $P_{in}=6000$ W, $p_{0}=30$ kPa. It can be clearly seen that with the increase in the pressure and power combinations, amplitude and area of the MWPD is increased around substrate stage. However, the change is not symmetric in $(r,z)$-directions. It also demonstrates that the microwave power coupling through co-axial mode transformer inside the cavity is good and satisfactory for the reactor operations. Since, MWPD is increasing, electron number density will also be increased as more power is available to be transferred to the neutral $H_{2}$ gas molecules via electron-$H_{2}$ chemical reactions listed in Table. \ref{TABLE_1A} . Fig. \ref{FIG_11_ELECTRON_DENSITY_VARIOUS_POWER_PRESURE} shows electron number density $n_{e}$ variations above the substrate stage for various power density parameters listed in Table \ref{TABLE 1} i.e (a) $P_{in}=3000$ W, $p_{0}=15$ kPa (b) $P_{in}=4000$ W, $p_{0}=20$ kPa (c) $P_{in}=5000$ W, $p_{0}=25$ kPa and (d) $P_{in}=6000$ W, $p_{0}=30$ kPa. As expected, we have observed that with the increase in the power and pressure electron number density also increases. Also, with the proper combination of power and pressure, similar to the MWPD variations $n_{e}$ profiles also does not show any symmetric change along $(r,z)$-directions and uniform density can be achieved.

\begin{figure*}
\centerline{\includegraphics[scale=0.45]{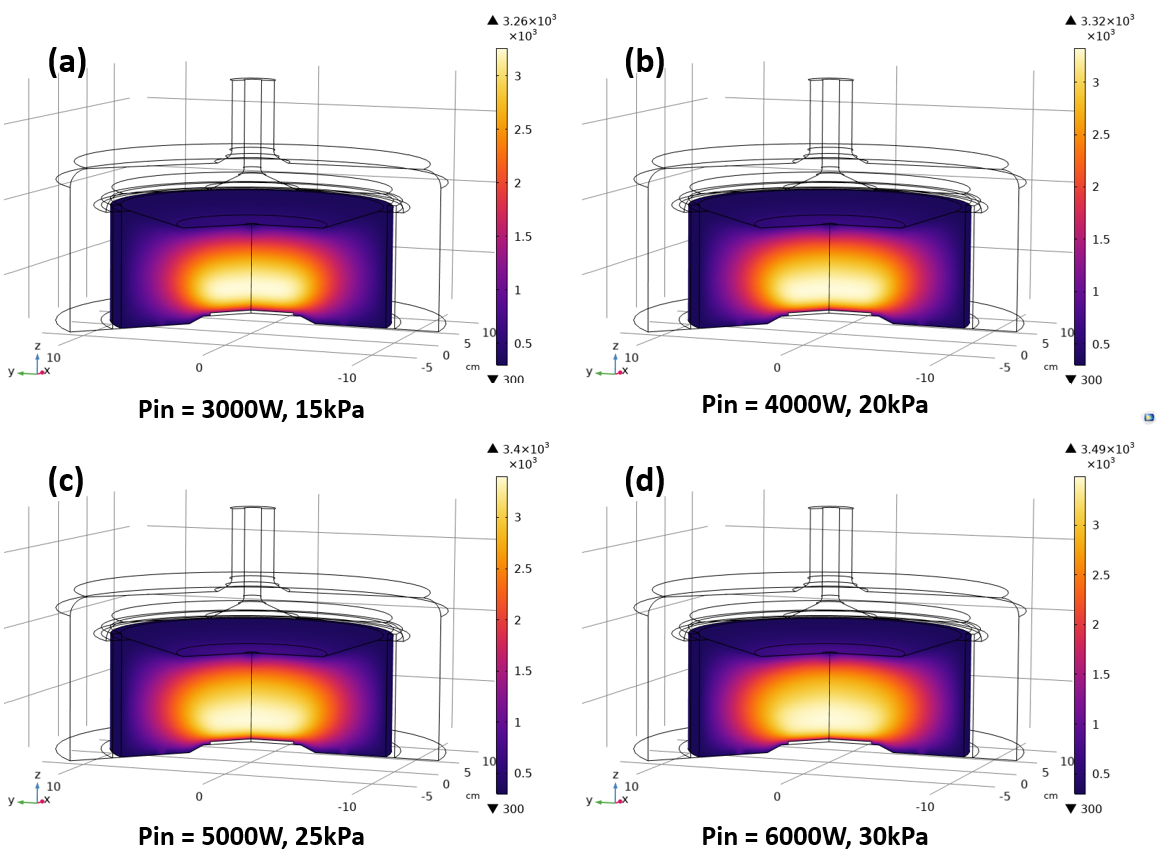}}
\caption{3D Contour variation of gas temperature $T_{g}$ around the substrate stage for different power and pressure parameters i.e (a)$P_{in}=3000$ W, $p_{0}=15$ kPa (b) $P_{in}=4000$ W, $p_{0}=20$ kPa (c) $P_{in}=5000$ W, $p_{0}=25$ kPa and (d) $P_{in}=6000$ W, $p_{0}=30$ kPa. One can observe that with the increase in power and pressure, gas temperature is also increased.}
\label{FIG_12_GAS_TEMPERATURE_VARIOUS_POWER_PRESURE}
\end{figure*}


\begin{figure*}
\centerline{\includegraphics[scale=0.30]{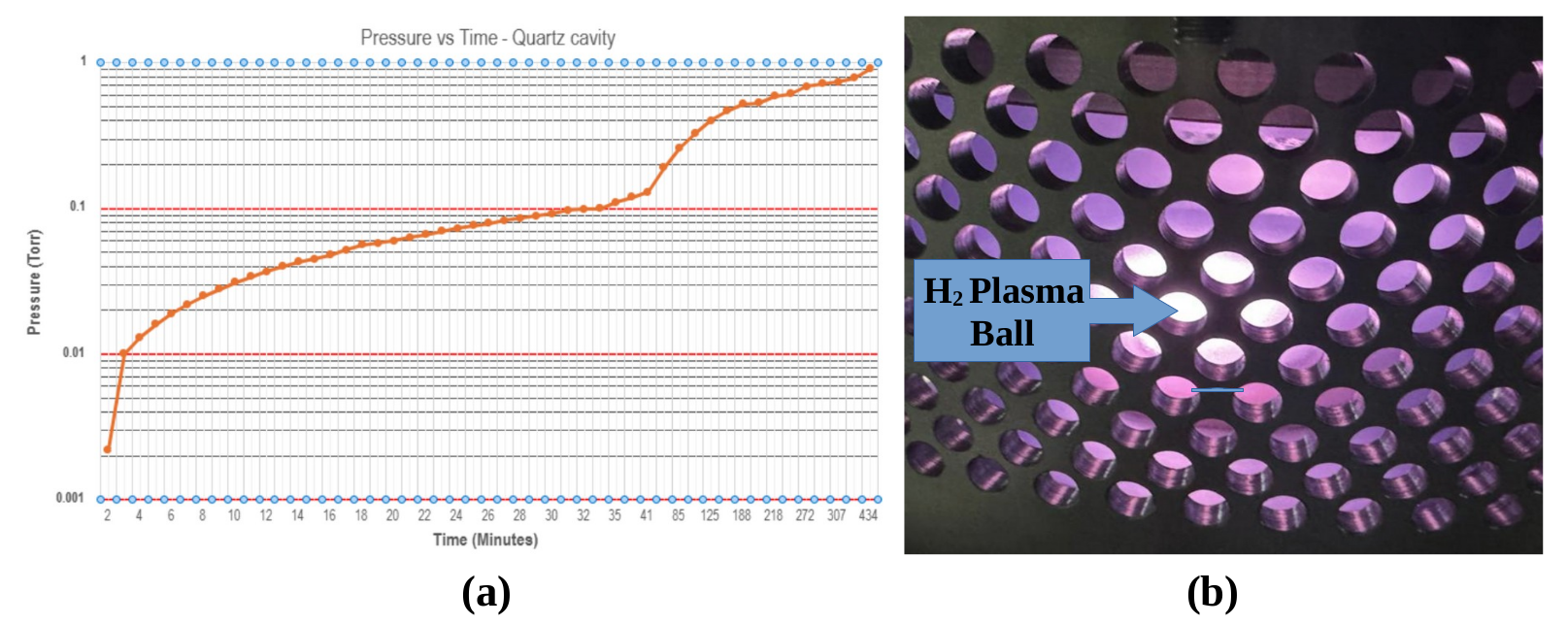}}
\caption{Experimental validation plot of the newly designed perforated cylindrical MPCVD reactor. In Fig. (a) we demonstrate the vacuum integrity of the quartz cavity by plotting pressure versus time variations. Fig.(b) shows actual image of the hydrogen ($H_{2}$) plasma discharge generated by a microwave power of $P_{in}=2$ kW and a gas pressure of $p_{0}=10$ kPa in the newly designed perforated cylindrical cavity.}
\label{FIG_14_Experimental_validation}
\end{figure*}

Fig. \ref{FIG_12_GAS_TEMPERATURE_VARIOUS_POWER_PRESURE} [(a), (b), (c) and (d)]demonstrates 3D contour variation of gas temperature $T_{g}$ above substrate stage, for various power and pressure parameters listed in Table \ref{TABLE 1} i.e (a)$P_{in}=3000$ W, $p_{0}=15$ kPa (b) $P_{in}=4000$ W, $p_{0}=20$ kPa (c) $P_{in}=5000$ W, $p_{0}=25$ kPa and (d) $P_{in}=6000$ W, $p_{0}=30$ kPa. With the proper increment in power and pressure, an increased rate of electron and neutral gas molecules collisions which leads to more energy transfer to the $H_{2}$ gas molecules, hence, we observe an expected increase in the gas temperature $T_{g}$ as shown in  our simulation results (Fig. \ref{FIG_12_GAS_TEMPERATURE_VARIOUS_POWER_PRESURE}). From Fig. \ref{FIG_12_GAS_TEMPERATURE_VARIOUS_POWER_PRESURE} (a) and (d), we can conclude that when the microwave power is simultaneously increased from $P_{in}=1000$ to $P_{in}=6000$ W alongwith gas pressure from $p_{0}=10$ to $p_{0}=30$ kPa, amplitude as well as region of effective high $T_{g}$ is also increased. However, Unlike individual variation in power, 2D $n_{e}$ profiles shown in Figs. \ref{FIG_8_2D_PLOT_PRESURE_CONST_POWER_VARY} and \ref{FIG_11_ELECTRON_DENSITY_VARIOUS_POWER_PRESURE}, $T_{g}$ profiles shows a symmetric increments along the radial (r) and axial (z) directions with combined increase in power and pressure. In the next Section, we present a couple of preliminary experimental validation tests performed on the newly designed and fabricated perforated MPCVD reactor.

\section{Experimental Validation}
\label{experimental_validation}

Validation of the design and simulation models are attained by performing experiments with the newly fabricated perforated MPCVD reactor. In this paper, we only present two major aspects of the experimental validation i.e. first, vacuum integrity check and second, hydrogen $H_{2}$ plasma discharge ignition. For vacuum check inside the cavity, we have made required connection with vacuum pump through butterfly valve and pirani gauge is connected for pressure measurement. The vacuum integrity test aims to detect any leaks or troubles in the vacuum cavity, allowing for modification and design changes in early stages and thereby guaranteeing reliable performance. Experimentally obtained Parameters used for this check is tabulated in Table. \ref{TABLE 2}. 

\begin{table}
\caption{Parameters for the vacuum integrity experiment on the novel MPCVD reactor cavity}  
\centering                         
\begin{tabular}{c c}           
\hline\hline                        
Parameters & Specifications   \\  [1.0ex] 
\hline          
Pumping speed of Edward dry & 16.7 L/s \\          
 Scroll pump &   \\          
Cavity volume to be pumped & 2.7 L \\
Time taken to reach $7 \times 10^{-3}$ torr & 7  mins \\ [1ex]
\hline\hline                               
\end{tabular}
\label{TABLE 2}
\end{table}

In Fig. \ref{FIG_14_Experimental_validation} [(a) and (b)], we present plots associated with the experimental validation of the newly designed cavity. In Fig. \ref{FIG_14_Experimental_validation} (a) we demonstrate the vacuum integrity of the quartz cavity by plotting pressure versus time variations. From the experiment, we have observed that degassing and the volume of the chamber play a crucial role in attaining higher vacuum integrity. The calculated leak rate is 26.5 sccm (0.72 torr L/s). Fig. \ref{FIG_14_Experimental_validation} (b) shows actual image of the hydrogen ($H_{2}$) plasma discharge generated at a microwave power of $P_{in}=2$ kW and a gas pressure of $p_{0}=10$ kPa in the newly designed perforated cylindrical reactor cavity. From the portrait, we can see that a plasma ball of pink color covers the surface of the substrate. The purple plasma ball was estimated to be around 50 mm in diameter, which is in good agreement with the theoretically predicted value. In the next section, we summarize and conclude the key results of the present work.


\section{Discussion and Conclusion}
\label{Discussion and conclusion}

In this work, we have presented an In-house development of new cylindrical microwave plasma $TM_{011}$ mode resonant cavity with perforations particularly used for diamond deposition process via MPCVD. We have followed the standard reactor designing methodology (Sec. \ref{new_reactor_design_prototype_assembly}), which includes electromagnetic (EM) simulations performed using high frequency Maxwell's solver of commercially available in COMSOL multiphysics software package. It gives us good idea of the electric field amplitude distribution inside the cavity. After EM simulations in vacuum conditions, we coupled plasma, flow and thermal modules to it which solves continuity, Navier-Stokes and heat conduction equations respectively, using radio frequency and plasma interfaces available in COMSOL multiphysics package. These simulation results illustrate various EM and plasma properties such as electron density $n_{e}$, electron temperature $T_{e}$, gas temperature $T_{g}$ in the plasma core as well as near substrate, electric field amplitude distribution, hydrogen number density $n_{H}$ which help us to optimize the reactor cavity design. Axi-symmetrical cylindrical geometry makes it easy for real time impedance tunning during reactor operations. In addition, for the reactor cavity, unique circumferential co-axial antenna alongwith cylindrical quartz were proposed as mode transformer and vacuum seal respectively (See Figs. \ref{FIG_2_Simulation_Domain_and_Schematic} and \ref{FIG_4_Prototype_Assembly}).

We have also investigated the effect of increase in pressure, power and power density amplitudes on hydrogen $H_{2}$ plasma discharge parameters i.e. $n_{e}$, $T_{e}$, $T_{g}$, electric field intensity, microwave power deposition (MWPD) and $n_{H}$ distributions inside the cavity above substrate. The important results are summarized as follows,
\begin{itemize}
    \item When we increased the gas pressure i.e. from $p_{0}=10$ kPa to $p_{0}=20$ kPa at constant microwave power $P_{in}=2$ kW, we observe an increase in electron density $n_{e}$, electric field inside the cavity, atomic hydrogen density $n_{H}$ respectively, whereas, increase in the gas temperature is noticed. Details of these observation are elaborately discussed in Sec. \ref{Effect_of_pressure_variation_on_plasma_characteristics}. 

    \item When we increased the microwave power i.e. from $P_{in}=2$ kW to $P_{in}=4$ kW at constant gas pressure $p_{0}=10$ kPa, we notice an expansion of the plasma region, an increase in microwave electric field, a slight reduction in electron density $n_{e}$, and gas temperature $T_{g}$, and inrease in atomic hydrogen density $n_{H}$ respectively. Details of these observation are elaborated in Sec. \ref{Effect_of_power_variation_on_plasma_characteristics}. 

    \item Upon simultaneous increase in the microwave power and gas pressure parameters tabulated in Table. \ref{TABLE 1}, we have observed,
    \begin{itemize}
        \item An symmetric and uniform increase in the $n_{e}$,  $T_{g}$ inside the plasma core as shown in Fig. \ref{FIG_9_1D_PLOT_Power_density_variations}.

        \item Order of $n_{e}$ and $n_{H}$ density profiles peaks around $10^{17}$ and $10^{22}~m^{-3}$ respectively.

        \item Gas temperature in the plasma core reaches above 3400 K at 6kW and 30kPa.

        \item Increase in 3D variations of microwave power deposition (MWPD), $n(e)$, and $T_{g}$ above the substrate or deposition area as shown in Figs. \ref{FIG_10_MWPD_VARIOUS_POWER_PRESURE}, \ref{FIG_11_ELECTRON_DENSITY_VARIOUS_POWER_PRESURE}, \ref{FIG_12_GAS_TEMPERATURE_VARIOUS_POWER_PRESURE}.

    \item Our simulation results indicate that the newly designed $TM_{011}$ reactor cavity can sustain high power density plasmas and it can operate upto a microwave power and gas pressure range of $P_{in}=6$ kW and $p_{0}=30$ kPa respectively. 
\end{itemize}
\end{itemize}

Our reactor cavity design methodology is further validated through hydrogen $H_{2}$ plasma discharge ignition experiment. Vacuum integrity tests have been performed and $H_{2}$ plasma was ignited with a microwave power and gas pressure of $P_{in}=2$ kW and $p_{0}=10$ kPa respectively. We have achieved a pink semi-spherical $H_{2}$ plasma ball of diameter around 50 mm which is in good agreement with the computationally predicted value. Further, long hours high power density experiments to test the plasma stability in our newly developed reactor cavity, reactor design modifications based on inputs from the long time experimental runs, experiments with inclusion of small percentage of gasses like methane $(CH_{4})$ and nitrogen $(N_{2})$ for diamond deposition and both theoretical as well as experimental diamond growth rate estimates in the newly designed perforated resonant cavity will be addressed in the near future.    

\section*{Acknowledgments}
Authors would like to acknowledge Indian Institute of Technology Madras and Ministry of Commerce and Industry (MOCI), New Delhi, India, for facilitating the establishment of “India Centre for Lab-Grown Diamond (InCent-LGD)”. [Grant : K-35011/11/2022-EP].

\section*{Data availability statement}
The data that support the findings of this study are available upon reasonable request from the authors.

\section*{References}
\bibliography{iopart-num}

\end{document}